\documentclass[review]{elsarticle}
\usepackage{graphicx}
\usepackage{subfigure}
\usepackage{algorithm, algorithmic}
\usepackage{lineno,hyperref}
\usepackage{amsmath}
\usepackage{diagbox}
\makeatletter
\renewcommand{\@thesubfigure}{\hskip\subfiglabelskip}
\makeatother
\newcommand{\tabincell}[2]{\begin{tabular}{@{}#1@{}}#2\end{tabular}}
\modulolinenumbers[5]

\journal{Journal of \LaTeX\ Templates}

\begin{document}

\begin{frontmatter}

\title{Emergence of Robust and Efficient Networks\\ in a Family of Attachment Models}
\tnotetext[mytitlenote]{Japan Advanced Institute of Science and Technology \href{http://www.jaist.ac.jp}.}


\author[mymainaddress]{Fuxuan Liao\corref{mycorrespondingauthor}}
\cortext[mycorrespondingauthor]{Corresponding author}

\author[mymainaddress]{Yukio Hayashi}


\address[mymainaddress]{Japan Advanced Institute of Science and Technology
, Nomi, Ishikawa, Japan}


\begin{abstract}
Self-organization of robust and efficient networks is important for the future designs of communication or transportation systems, because both characteristics are not coexisting in many real networks. As one of the candidates for the coexisting, the optimal robustness of onion-like structure with positive degree-degree correlations has recently been found, and it can be generated by incrementally growing methods based on a pair of random and intermediation attachments with the minimum degree selection. In this paper, we introduce a continuous interpolation by a parameter $\beta\geq 0$ between random and the minimum degree attachments to investigate the reason why the minimum degree selection is important. However, we find that the special case of the minimum degree attachment can generate highly robust networks but with low efficiency as a chain structure. Furthermore, we consider two intermediation models modified with the inverse preferential attachment for investigating the effect of distance on the emergence of robust onion-like structure. The inverse preferential attachments in a class of mixed attachment and two intermediation models are effective for the emergence of robust onion-like structure. However, a small amount of random attachment is necessary for the network efficiency, when $\beta$ is large enough. Such attachment models indicate a prospective direction to the future growth of our network infrastructures.

\end{abstract}


\end{frontmatter}


\section{Introduction}
Unfortunately, a common topological structure called scale-free (SF) in many social, technological, and biological networks \cite{barabasi1999emergence} is extremely vulnerable against intentional attacks to nodes of hubs with huge degrees \cite{albert2000error}, while it has efficient short paths between two nodes. The combination of efficient paths and extreme vulnerability makes a double-edged sword. Moreover, the SF network can be generated by a selfish rule of the preferential attachment, such as in Barab{\'a}si and Albert (BA) model \cite{barabasi1999mean}, whose linked node from a new node is chosen with probability proportional to its degree. However, we aim to find unselfish better attachments for equipping both important properties of strong robustness of connectivity and high network efficiency than the conventional ones \cite{barabasi1999emergence,albert2000error,barabasi1999mean} in growing networks.\\

Over the past few decades, there is a variety of attachments in growing networks. For instance, a mixed attachment between random and preferential attachments has been discussed \cite{Shao_2006,DegreeDistribution2012}. Other mixed attachment between preferential attachment and triad formation for tunable clustering \cite{holme2002growing} has also been studied. In a growing network model \cite{krapivsky2000connectivity,krapivsky2001organization,krapivsky2002statistical}, a node with degree $k$ is selected as the link destination with probability proportional to $k^{\upsilon}$, $\upsilon \ge 0$. The weak and strong preferential attachments continuously change degree distributions from exponential, power-law with an exponential cut-off, power-law as SF, and to the exclusive star-like network according to the parameter value: $\upsilon=0$, $0<\upsilon<1$, $\upsilon=1$, and $\upsilon>1$, respectively. Moreover, duplication \cite{pastor2003evolving} or copying \cite{PhysRevE.71.036118} by attachment to the nearest neighbors of a randomly chosen node generates a power-law distribution derived in approximate analyses. However, the inverse preferential attachment has been hardly considered except for a special case of $\beta=1$ \cite{e22091029,Siew2020}. In particular, beyond the analysis of degree distributions \cite{zadorozhnyi2015growing}, the properties of network efficiency and robustness of connectivity have not been discussed for the inverse preferential attachment. This paper reveals these important properties for the $k^{-\beta}$-attachment in varying $\beta$ values.\\

On the other hand, it has been found that the onion-like topological structures with positive degree-degree correlations gives the optimal robustness against intentional attacks under a given degree distribution \cite{schneider2011mitigation,PhysRevE.85.046109}. Such structure is visualized, when similar degree nodes locate on concentric circles in decreasing order of degrees from core to peripheral. Some rewiring methods have been proposed to generate onion-like networks \cite{schneider2011mitigation,PhysRevE.84.026106}. However, these methods discard already existing links. Therefore, they are difficult to be applied for improving the robustness in real networks. While incremental growing methods have also been proposed by applying cooperative partial copying with adding shortcuts \cite{hayashi2014growing} or intermediation \cite{hayashi2018new,hayashi2018onion}. Moreover, in the onion-like networks, both robustness and efficiency coexist \cite{hayashi2018new}. The connection between randomly chosen and a distant node through intermediation \cite{hayashi2018new,hayashi2018onion} has been inspired from long-distance relations in case studies in organization theory: long-distance relations contribute to overcoming the crisis of Toyota group's supply chain damaged by a large fire accident to their subcontract plants \cite{nishiguchi1998toyota,nishiguchi2000fractal,nishiguchi2007}, and to rapidly organizing world-wide economic networks with expanding business chances by Wenzhou people in China \cite{nishiguchi2007}. However, it has not been comprehensively understood which of random, intermediation, and inverse preference attachments are dominant for the emergence of onion-like networks.\\

The organization of this paper is as follows. In Section 2, we introduce the parametric interpolation as the $k^{-\beta}$-attachment between the random attachment and the minimum degree attachment at $\beta=0$ and $\beta \to \infty$. In Subsection 2.2, we numerically investigate a condition of the network efficiency whose average length is logarithmic order of the node size $N$ in a growinhttps://arxiv.org/userg network by the mixed attachment between the random attachment and the $k^{-\beta}$-attachment. In Section 3, to investigate the effects of distance and the minimum degree selection, we slightly modify similar models of propinquity \cite{gallos2019propinquity} and intermediation \cite{hayashi2018new,hayashi2018onion} with the $k^{-\beta}$-attachment, and show that the $k^{-\beta}$-attachment is dominant for the emergence of onion-like networks. We summarize these results in Section 4. The numerical results are given by the average values over 100 realizations of randomly generated networks.  



\section{Emergence of Onion-like Networks with Efficient Paths}
The following processes are common in growing networks by attachments of links. At each time step, a new node is added and connects to existing $m$ nodes by attachments in prohibiting self-loops and multi-links, as similar to BA model \cite{barabasi1999mean}. However, the selections of linked nodes are different in a family of attachment models introduced in the next subsection.

\subsection{Dominance of the Minimum Degree Attachment for Onion-like Networks}
We focus on random and the minimum degree selections of nodes, because they are necessary for generating a strongly robust onion-like networks as follows. In order to make onion-like networks with positive assortativity as degree-degree correlations \cite{PhysRevE.85.046109}, incrementally growing methods based on a pair of random and intermediation (MED) attachments have been proposed \cite{hayashi2018new,hayashi2018onion}. With either the minimum degree or random selection, the type in MED model is called MED-kmin or MED-rand \cite{hayashi2018onion}, respectively. In each pair of MED-kmin, one of link destination is chosen uniformly at random (u.a.r), and the other link destination is a node with the minimum degree for intermediation in $\mu$ hops from the randomly chosen pair node. Intermediation in $\mu$ hops means an attachment to a node in the $\mu+1$-th neighbors. Moreover, through numerical simulations, by the iterative attachments to a node at $\mu \geq 3,4$ hops, relatively long loops are formed and enhance the robustness \cite{gallos2019propinquity}. If the network has a small amount of loops, it can become a tree by removing nodes. Therefore, it is easily fragmented by further attacks to the articular nodes. This phenomenon is supported by the equivalence of dismantling and decycling problems\cite{hayashi2018onion}. Similarly in MED-rand, instead of the nodes with the minimum degree, a node is chosen u.a.r in the $\mu+1$-th neighbors from the randomly chosen pair node. Since older nodes with large degrees tend to be selected in the random attachment, this first attachment contributes to enhancing positive correlations among large degree nodes. The other attachment to the nodes with the minimum degree contributes to enhancing positive correlations among small degree nodes. In other words, the second attachment establishes a connection between the minimum degree node in the $\mu+1$-th neighbors and a new node of the minimum degree $m$ in the whole network, therefore it enhances positive correlations among small degree nodes. Note that onion-like networks emerge by MED-kmin, but not by MED-rand \cite{hayashi2018onion} because of a weak effect of the second attachment. A modification of MED-kmin will be discussed including the effect of distance regulated by a parameter $\mu$ in Section 3.\\

In the remaining of this subsection, we show that onion-like networks can also emerge by $k^{-\beta}$-attachment, $0<\beta<\infty$, as an interpolation of random and minimum degree attachments at $\beta=0$ and $\beta \to \infty$. Here, no effect of distance is included in the attachments until the related discussion to MED in Section 3. First, we derive a degree distribution, in which the maximum degree is bounded as $\beta$ becomes larger. Second, for the minimum degree attachment at $\beta \to \infty$, we find that a special chain structure is obtained, the network efficiency is low because of the long distance paths. Third, we numerically investigate three measures: assortativity as degree-degree correlations, robustness of connectivity, and network efficiency in order to reveal which values of $\beta$ contribute to generating both efficient and robust onion-like networks by the $k^{-\beta}$-attachment.\\

\begin{figure}[htbp]
\centering
\subfigure[(a) $m=2$, $\beta=1$]{
\begin{minipage}[t]{0.5\linewidth}
\centering
\includegraphics[width=60mm]{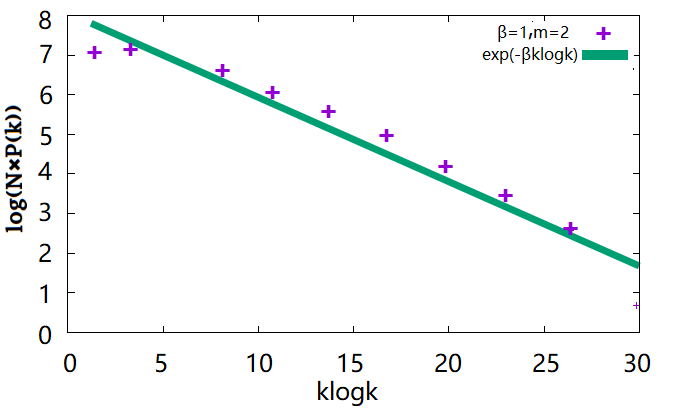}
\end{minipage}%
}%
\subfigure[(d) $m=4$, $\beta=1$]{
\begin{minipage}[t]{0.5\linewidth}
\centering
\includegraphics[width=60mm]{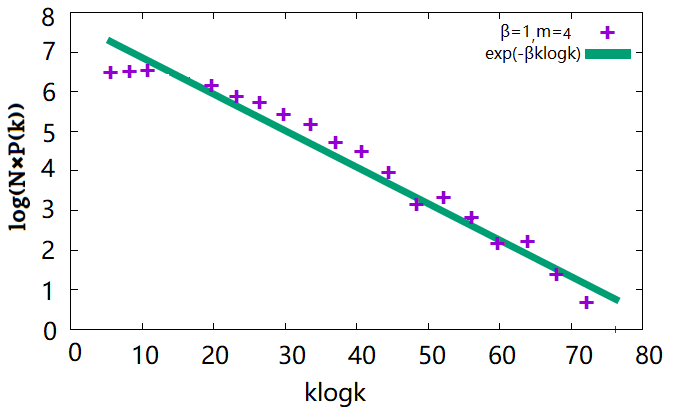}
\end{minipage}%
}%

\subfigure[(b) $m=2$, $\beta=4$]{
\begin{minipage}[t]{0.5\linewidth}
\centering
\includegraphics[width=60mm]{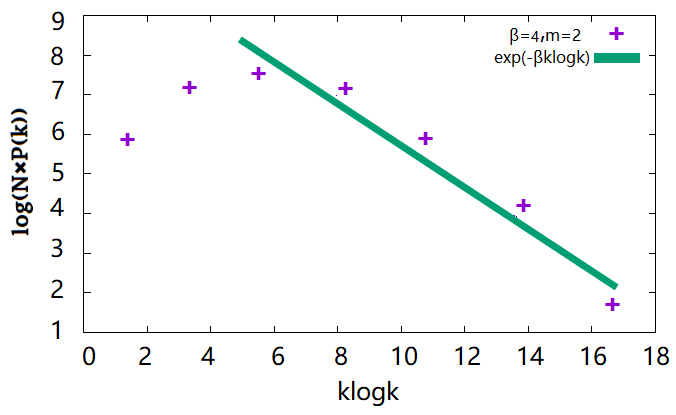}
\end{minipage}%
}%
\subfigure[(e) $m=4$, $\beta=4$]{
\begin{minipage}[t]{0.5\linewidth}
\centering
\includegraphics[width=60mm]{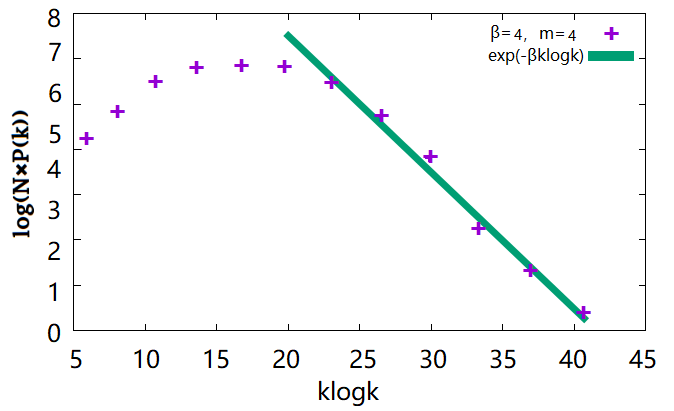}
\end{minipage}%
}%

\subfigure[(c) $m=2$, $\beta=10$]{
\begin{minipage}[t]{0.5\linewidth}
\centering
\includegraphics[width=60mm]{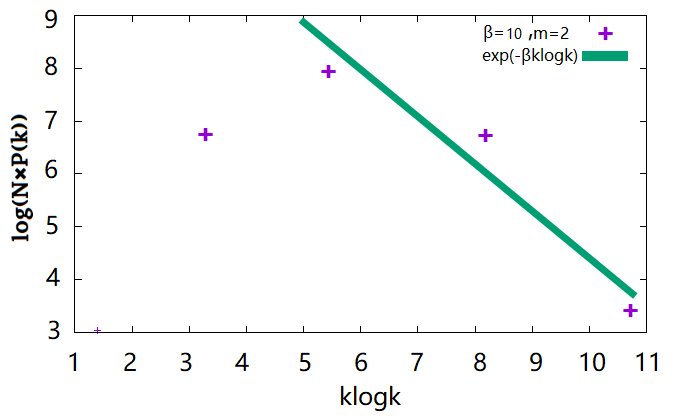}
\end{minipage}%
}%
\subfigure[(f) $m=4$, $\beta=10$]{
\begin{minipage}[t]{0.5\linewidth}
\centering
\includegraphics[width=60mm]{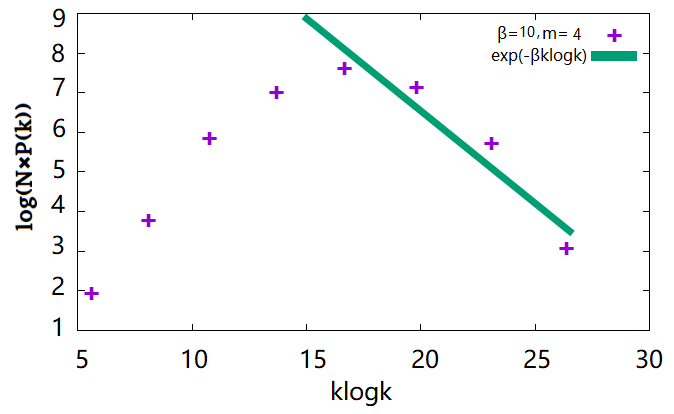}
\end{minipage}%
}%
\centering
\caption{Fitting the tail to the estimated $e^{-\beta klogk}$ as a green line. Purple points and green lines denote the degree distribution and the estimated one in the growing networks by the $k^{-\beta}$-attachment. The number of adding links per time step is $m=2$ (left) and $m=4$ (right). As the parameter value of $\beta$ is changed from small to large in (a) (d) $\beta=1$, (b) (e) $\beta=4$, and (c) (f) $\beta=10$, the slope becomes steeper. Note that each range of the horizontal axis is different in (a)$\sim$(f).} \label{fig1}
\end{figure}

First, as a parametric interpolation between random and minimum degree attachments (at $\beta=0,\infty$), we consider the $k^{-\beta}$-attachment to an existing node $i$ chosen with probability $\frac{k_{i}^{-\beta}}{\sum_{j}k_{j}^{-\beta}}$, $0 < \beta < \infty$, for adding $m$ links from a new node at each time step. In general, for a nonlinear attachment to a node $i$ chosen with probability function $f(k_{i})$ of its degree $k_{i}$, the degree distribution $P(k)$ is derived from Eq.(9) in Ref.\cite{zadorozhnyi2015growing} as follows.
\begin{eqnarray*}
P(k) &=& \frac{\langle f \rangle}{\langle f \rangle+m\times f(K_{\rm min})} \prod_{\kappa=K_{\rm min}+1}^{k} \frac{m\times f(\kappa-1)}{\langle f \rangle+m \times f(\kappa)}, \\
\langle f \rangle &\overset{def}{=}&\sum_{k=K_{\rm min}}^{K_{\rm max}} f(k)\times P(k), \qquad (2.1)
\end{eqnarray*}
where $K_{\rm min}$ and $K_{\rm max}$ denote the minimum and maximum degrees in a network. In the case of $f(k)= k^{-\beta}$ for $m=K_{\rm min}\leq k\leq K_{\rm max}<\infty$, it is rewritten as,
\begin{eqnarray*}
P(k) &=& \mathrm{exp} \left[ \mathrm{log}\left\{\frac{\langle f \rangle}{\langle f \rangle+m\times f(K_{\rm min})} \prod_{\kappa=K_{\rm min}+1}^{k} \frac{m\times f(\kappa-1)}{\langle f \rangle+m \times f(\kappa)} \right\}\right], \\
&=& \mathrm{exp}\left[ \sum_{\kappa=K_{\rm min}+1}^{k}  \left\{ \mathrm{log}(m)+\mathrm{log}((\kappa-1)^{-\beta})-\mathrm{log}(\langle f \rangle+m\times \kappa^{-\beta})\right\}+C\right], \qquad (2.2)
\end{eqnarray*}
where $C=\mathrm{log}\frac{\langle f \rangle}{\langle f \rangle+m\times f(K_{\rm min})}$ is a constant. In the right-hand side of Eq.(2.2), the constant terms $C$, $\mathrm{log}(m)$, and $\mathrm{log}(\langle f \rangle+m\times K_{\rm min}^{-\beta})>\mathrm{log}(\langle f \rangle+m\times \kappa^{-\beta})$, $K_{\rm min}+1\leq \kappa \leq k$, are not dominant and ignored. Thus, it is approximated as the order of $\mathrm{exp}\left[ \int -\beta \mathrm{log}(\kappa) d\kappa\right] \sim e^{-\beta k\mathrm{log}k}$ without the constant factors. Fig. \ref{fig1} shows a good fitting for the above estimation at $m=2,4$ and $\beta=1,4,10$ by the least squares method for $\mathrm{log}(N\times P(k))$ and $k\mathrm{log}k$. We remark that the maximum degree is decreasing as larger $\beta$ in Fig. \ref{fig1} and bounded by $2m$ for $\beta \to \infty$ as mentioned later. Note that there is a monotonic increasing relation between $k$ and $k\mathrm{log}k$, $k\geq1$. 

\begin{figure}[htbp]
\centerline{\includegraphics[clip,width=130mm]{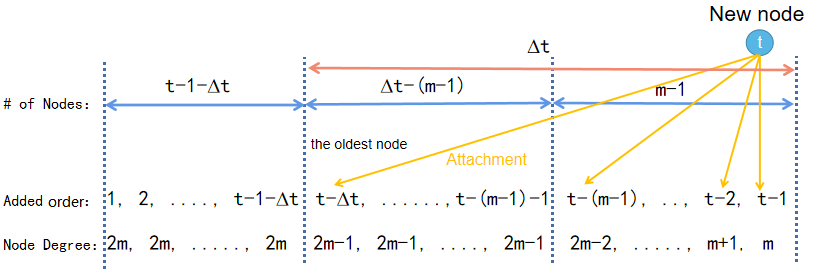}}
\caption{Link destinations by the minimum degree attachment at $\beta \to \infty$. black lines show the attachment from a new node to existing $m$ nodes by the minimum degree attachment. There are three parts from right to left. In the first part (right) of $m-1$ nodes, a new node connects to the existing $m-1$ nodes whose degree are the smallest $m, m+1 ..., 2m-2$, because of the minimum degree attachment. In the second part (middle) of $\Delta t - (m-1)$ nodes, the remaining one link connects to a node with degree $2m-1$. Since there are $\Delta t$ nodes whose degree is less than $2m$, the added time of the oldest node with $2m-1$ degree is $t-\Delta t$. In the third part (left) of $t-1-\Delta t$ nodes, there are $t-1-\Delta t$ nodes with degree $2m$. The relation of the added ordering of these nodes and their degrees at time t is shown as first and second rows from the bottom.}\label{fig2}
\end{figure}

Second, we reveal a special chain structure in the case of $\beta \to \infty$ as the minimum degree attachment. By the attachment at time step $t$, a new node connects to the existing nodes added at $t-1$, $t-2$, ..., $t-(m-1)$, whose degree are the smallest $m, ..., 2m-2$. Because the increase of degree is only one at each time step in prohibiting multi-links. In the total $m$ links, the remaining one link connects to a node with degree $2m-1$ as the next smaller than $2m-2$ because of the minimum degree attachment. Fig. \ref{fig2} shows a special case of selection to the oldest node with degree $2m-1$ in order to simplify the discussion. While nodes with degree $2m$ do not get a link anymore, since there exists more than one node with degree $2m-1$. We confirm it from the above discussion and Fig. \ref{fig2} as follows. For the sum $M_{t}$ of degrees at time step $t$, the following equation holds.
\begin{eqnarray*}
M_{t} &=& 2M_{0}+2mt, \\
&=& 2m(N_{0}+t-\Delta t)+(2m-1)(\Delta t-m) \\
&& +\frac{((m+1)+(2m-2+1))(m-1)}{2}+m,  \qquad (2.3)
\end{eqnarray*}
where $M_{0}$ and $N_{0}$ are the total numbers of links and nodes in the initial network at $t=0$. $\Delta t$ represents the difference of node IDs defined by the current and the insert time of the oldest node whose degree is less than $2m$. Therefore, $\Delta t$ is not the time duration but the number of nodes. In the right-hand side of Eq.(2.3), $2m(N_{0}+t-\Delta t)$ and $(2m-1)(\Delta t-m)$ denote the sums of degrees of $2m$ and $2m-1$. The third term $\frac{((m+1)+(2m-2+1))(m-1)}{2}$ is the sum of arithmetic sequence from $m+1$ to $2m-2+1$. The last term is the degree $m$ of a new node. We solve Eq.(2.3), 
\begin{eqnarray*}
\Delta t &=& 2mN_{0}-2M_{0}-\frac{m(m-1)}{2}. \qquad (2.4)
\end{eqnarray*}
For the initial complete graph, $N_{0}=m+1$ and $M_{0}=\frac{m(m+1)}{2}$ are substituted into Eq.(2.4). Then, we obtain the constant number
\begin{eqnarray*}
\Delta t = \frac{m(m+3)}{2}>m. \qquad (2.5)
\end{eqnarray*}
From Eq.(2.5), the number of nodes with $2m-1$ is $\Delta t-(m-1)>1$, therefore more than one node with degree $2m-1$ exists.\\

We remark that the network generated by the minimum degree attachment is an almost regular graph, since many $t-1-\Delta t$ nodes have degree $2m$ after a long time. Moreover, we emphasize that the network has a chain structure with interval of $\Delta t$. In the chain, the longest length of all the shortest paths (the diameter of network) is estimated as $N/\Delta t$. Tables 1 and 2 show that the practical measurement is very close to the estimated value $N/\Delta t$. The slight differences are due to that a node with degree $2m-1$ is randomly selected in the measurement instead of the oldest one in the estimation for a special case of Fig. \ref{fig2}. For $0<\beta<\infty$, the length is longer as $\beta$ increases. Note that the case of $\beta$=0 is random attachment.

\begin{table}[htbp]
\centering
\begin{tabular}{c|c|c|c}
m & Practical & Estimated $N/\Delta t$ & $\Delta t$ \\ \hline
2 & 1001 & 1000 & 5 \\ \hline
4 & 365 & 357 & 14 \\ \hline
\end{tabular}
\caption{Practical measurement of the diameter and the estimated $N/\Delta t$ for $N=5000$ generated by the minimum degree attachment with $\beta\to\infty$ from the initial complete graph with $m+1$ nodes for $m=2,4$.}\label{table1}
\end{table}

\begin{table}[htbp]
\centering
\begin{tabular}{c|c|c|c|c|c}
\diagbox[dir=SE]{m}{$\beta$} & 0 & 1 & 10 & 50 & 100 \\ \hline
2 & 10 & 11 & 12 & 285.8 & 727.5 \\ \hline
4 & 7 & 7 & 7 & 13.5 & 103 \\ \hline
\end{tabular}
\caption{Practical measurement of the diameter generated by the $k^{-\beta}$-attachment with several values of $\beta$ from the initial complete graph with $m+1$ nodes. The cases of $m=4$ at the bottom are shorter than ones of $m=2$ at the top. In addition, the path length also increases as $\beta$ increases.}\label{table2}
\end{table}

\begin{figure}[htbp]
\centering
\subfigure[(a) Assortativity]{
\begin{minipage}[t]{0.5\linewidth}
\centering
\includegraphics[width=50mm]{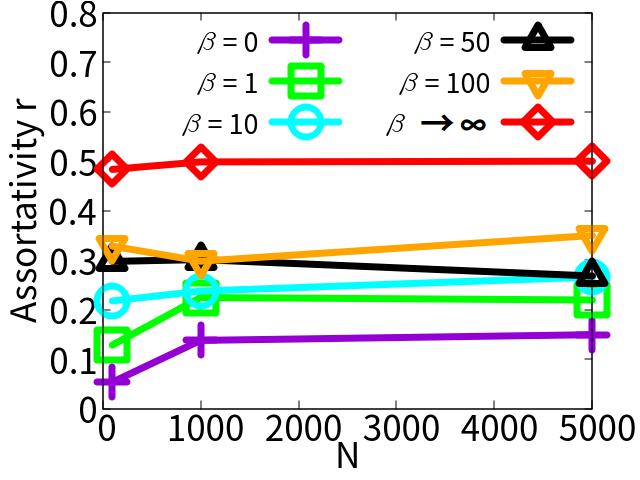}
\end{minipage}%
}%
\subfigure[(d) Assortativity]{
\begin{minipage}[t]{0.5\linewidth}
\centering
\includegraphics[width=50mm]{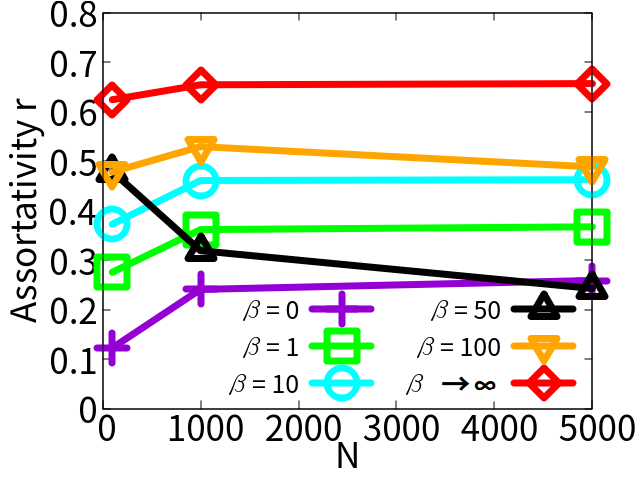}
\end{minipage}%
}%

\subfigure[(b) Robustness]{
\begin{minipage}[t]{0.5\linewidth}
\centering
\includegraphics[width=50mm]{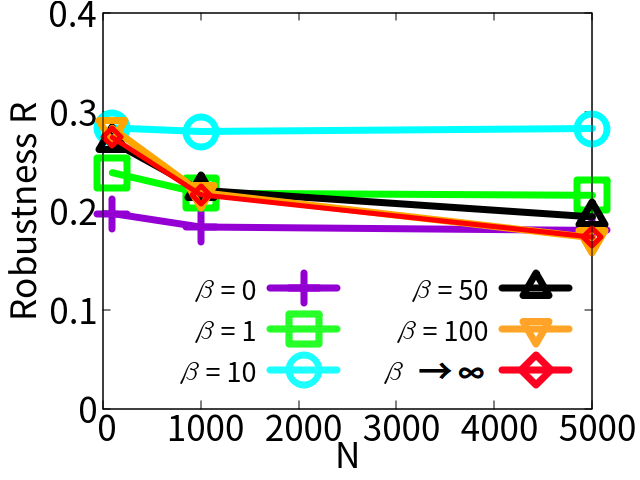}
\end{minipage}%
}%
\subfigure[(e) Robustness]{
\begin{minipage}[t]{0.5\linewidth}
\centering
\includegraphics[width=50mm]{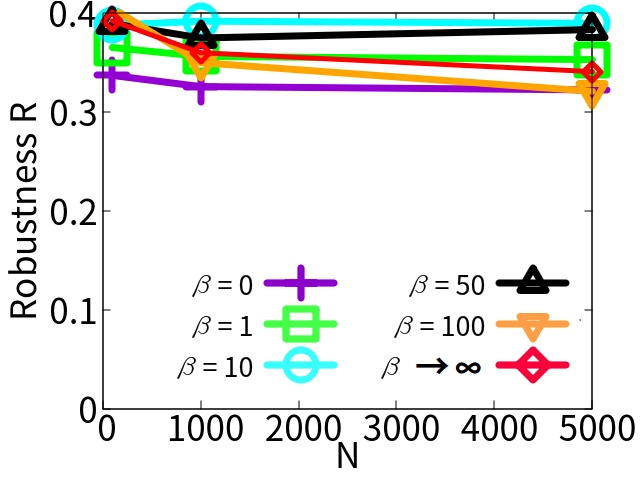}
\end{minipage}%
}%

\subfigure[(c) Efficiency]{
\begin{minipage}[t]{0.5\linewidth}
\centering
\includegraphics[width=50mm]{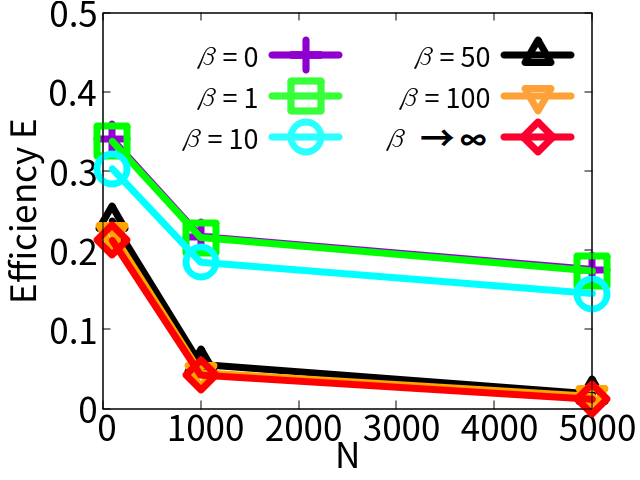}
\end{minipage}%
}%
\subfigure[(f) Efficiency]{
\begin{minipage}[t]{0.5\linewidth}
\centering
\includegraphics[width=50mm]{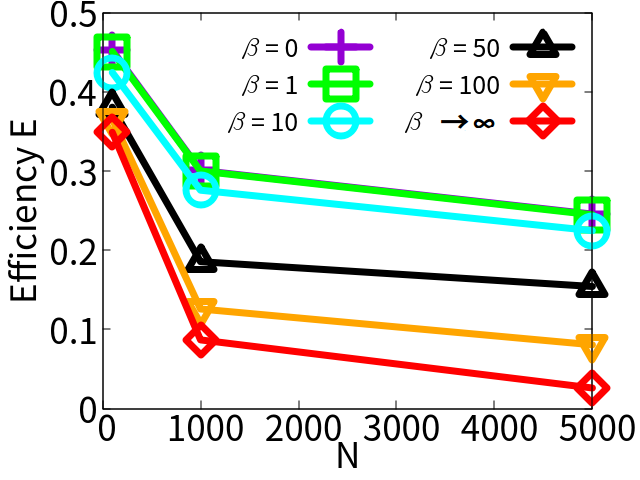}
\end{minipage}%
}%
\centering
\caption{Three measures at Top:$r$, Medium:$R$, Bottom:$E$ versus node size $N$ in the network by the $k^{-\beta}$-attachment with $m$=2 (left) and 4 (right) links per time step. Purple (cross mark), green (square mark), cyan (circle mark), black (triangle mark), and orange (inverse triangle mark) lines correspond to $\beta$ = 0, 1, 10, 50, and 100. Red line (diamond mark) indicates stronger assortativity than other color lines in (a) (d). The robustness for $m=4$ in (e) are higher than that in (b) for $m=2$. The case of red line (diamond mark) for $m=4$ shows that the network has both high assortativity and robustness as onion-like networks. The efficiency for $\beta \to \infty$ becomes lower than ones for $\beta$ = 0, 1, 10, 50, and 100 colored by purple, green, cyan, black and orange in (c) (f), because of a chain structure as mentioned in Fig. \ref{fig2} and Tables 1 and 2. Thus, the networks generated by the $k^{-\beta}$-attachment with large $\beta$ have stronger assortativity and robustness but weak efficiency. }\label{fig3}
\end{figure}

To reveal which values of $\beta$ are contributing for generating both efficient and robust onion-like networks by the $k^{-\beta}$-attachment, we investigate the following three measures: assortativity as degree-degree correlations \cite{newman2002assortative,newman2010epidemics}, robustness of connectivity \cite{schneider2011mitigation}, and network efficiency \cite{latora2001efficient}. Because the onion-like networks have high assortativity and robustness.
\begin{eqnarray*}
\mbox{Assortativity:} \quad r=\frac{S_{1}S_{e}-S_{2}^{2}}{S_{1}S_{3}-S_{2}^{2}},\qquad (2.6)
\end{eqnarray*}
\begin{eqnarray*}
\mbox{Robustness index:} \quad R=\sum_{Nq=1}^N\frac{S(Nq)}{N},\qquad (2.7)
\end{eqnarray*}
\begin{eqnarray*}
\mbox{Efficiency:} \quad E=\frac{1}{N(N-1)}\sum_{i\ne j}\frac{1}{L_{ij}}, \qquad (2.8)
\end{eqnarray*}
where $S_1=\sum_ik_i$, $S_2=\sum_ik_i^2$, $S_3=\sum_ik_i^3$, $S_e=\sum_{ij}A_{ij}k_ik_j$, and $A_{ij}$ denotes the $ij$ element of the adjacency matrix. $k_i$ and $k_j$ denote degrees at end-nodes of $(i,j)$ link. The positive or negative assortativity is distinguished by the sign $r>0$ or $r< 0$. $S(Nq)$ denotes the number of nodes included in the giant component (GC as the largest connected cluster) after removing $Nq$ nodes, where $Nq$ is the number of removed nodes by attacks with recalculation of the highest degree node as the target. $1/E$ is the harmonic mean of the shortest path length $L_{ij}$. $L_{ij}$ is counted by the number of hops between nodes $i$ and $j$.\\

Fig. \ref{fig3} (a) and (d) show the assortativity for $\beta=0,1,10,50,100$, and $\beta \to \infty$ in the growing networks by the $k^{-\beta}$-attachment. Red line with diamond mark ($\beta \to \infty$) indicates stronger assortativity than purple with cross mark ($\beta=0$), green with square mark ($\beta=1$), cyan with circle mark ($\beta=10$), black with triangle mark ($\beta=50$), and orange with inverse triangle mark ($\beta=100$) lines in Fig. \ref{fig3} (a) (d). Therefore, the case of larger $\beta$ has higher assortativity. In comparison with the same color lines or marks, the assortativity in Fig. \ref{fig3} (d) for $m=4$ are higher than that in Fig. \ref{fig3} (a) for $m=2$. Moreover, as shown in Fig. \ref{fig3} (b) (e), the robustness in Fig. \ref{fig3} (e) for $m=4$ are higher than that in Fig. \ref{fig3} (b) for $m=2$ in comparison with the same color lines or marks. In addition, the lines of assortativity $r$ increase monotonically for varying $\beta$ from 0 to $\infty$ in Fig. \ref{fig3} (a), while they increase for varying $\beta$ from 0 to 10 in Fig. \ref{fig3} (b), decrease for varying $\beta$ from 10 to 50, and increase again for varying $\beta$ from 50 to $\infty$. Moreover, the lines of robustness $R$ increase for varying $\beta$ from 0 to 10, decrease for varying $\beta$ from 10 to 50, and increase again for varying $\beta$ from 50 to 100 in Fig. \ref{fig3} (b) (e). After $\beta$ is greater than 100, the robustness approach to a constant $R$ value. The lines of efficiency $E$ decrease monotonically for varying $\beta$ from 0 to $\infty$  in Fig. \ref{fig3} (c) (f). The reason why assortativity $r$ and robustness $R$ change in this way is unknown exactly. However, a chain structure may affect them as discussed previously. In particular, red lines at $\beta \to \infty$ in Fig. \ref{fig3} (d) (e) for $m=4$ show that the network has both high assortativity ($r>0.2$) and robustness ($R>0.3$) as onion-like networks. Note that too strong assortativity lead to rather weaken the robustness with decreasing the value of $R$ \cite{PhysRevE.85.046109}. Remember that there are no exact conditions for the values of $r$ and $R$ to be an onion-like network, we consider them as $r>0.2$ and $R>0.3$, because BA model is not an onion-like, which is $r\approx 0$ and $R<0.23$ for the same $N$ and $m$ \cite{hayashi2018new,hayashi2018onion}. Fig. \ref{fig3} (c) (f) show that the efficiency for $\beta \to \infty$ (red line) becomes lower than ones for $\beta$ = 0, 1, 10, 50, and 100 (purple, green, cyan, black, and orange lines). The reason of low efficiency is due to a chain structure as mentioned in Fig. \ref{fig2} and Tables 1 and 2.

\subsection{Necessary Small Amount of Random Attachment for Efficient Paths}
We investigate the effect of random attachment on the network efficiency whose average lengths are $O(\mathrm{log}N)$ in the mixed attachment of the random attachment and the $k^{-\beta}$-attachment. It is worth to mention that the effect is nontrivial in growing networks, especially by the inverse preferential attachments for a large $\beta$. For example, the diameters are similar as 725.5 and 1000 (365 and 103) obtained by the practice at $\beta=100$ and the estimation for $m=2$ ($m=4$). When $\beta$ is large enough, the diameter becomes $O(N)$ because of approaching to the chain structure. In the following mixed attachment, it is the same that a new node is added at each time step, but the link destinations are chosen by random attachment with probability $p$ and by the $k^{-\beta}$-attachment with probability $1-p$. Note that the case of $\beta=0$ is pure random attachment.\\

\begin{figure}[]
\subfigure[(a) $m=2$]{
\begin{minipage}[t]{0.5\linewidth}
\centering
\includegraphics[width=60mm]{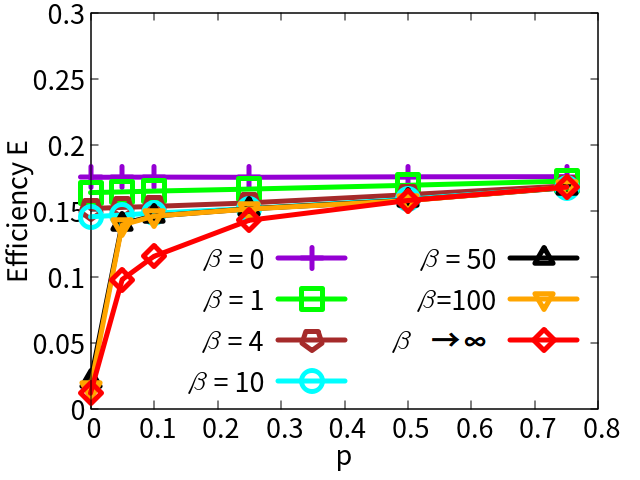}
\end{minipage}%
}%
\subfigure[(b) $m=4$]{
\begin{minipage}[t]{0.5\linewidth}
\centering
\includegraphics[width=60mm]{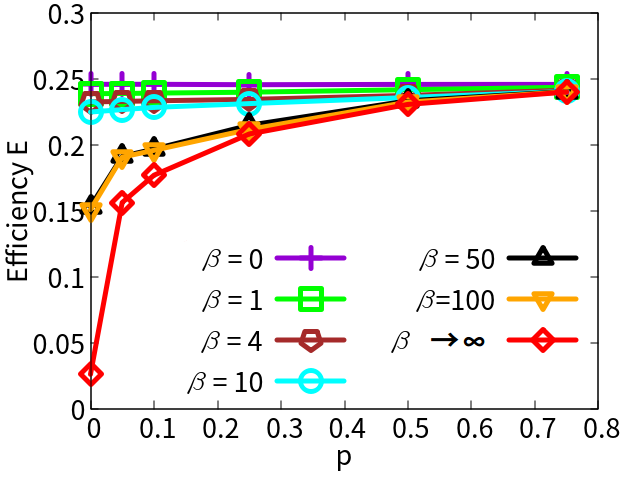}
\end{minipage}%
}%
\centering
\caption{Efficiency versus the mixing probability $p$ in the network by the mixed attachment of random and the $k^{-\beta}$-attachments with $m$=2 (left) and 4 (right) at node size $N=5000$. Here $p$ indicates the ratio of random attachment. Purple (cross mark), green (square mark), brown (pentagon mark), cyan (circle mark), black (triangle mark), orange (inverse triangle mark), and red (diamond mark) lines correspond to $\beta=0,1,4,10,50,100$, and $\beta \to \infty$. Green (square mark), brown (pentagon mark), and cyan (circle mark) lines show that the efficiency trends to slightly increase as the ratio $p$ is larger, and the efficiency rapidly increases for the horizontal axis as shown by the black (triangle mark), orange (inverse triangle mark), and red (diamond mark) lines. In comparison with the same color lines or marks, the efficiency in (b) for $m=4$ is higher than that in (a) for $m=2$ by using more adding links in the emergence of onion-like networks.}\label{fig5}
\end{figure}

\begin{figure}[htbp]
\centering
\subfigure[(a) $p=0$, $m=2$]{
\begin{minipage}[t]{0.5\linewidth}
\centering
\includegraphics[width=60mm]{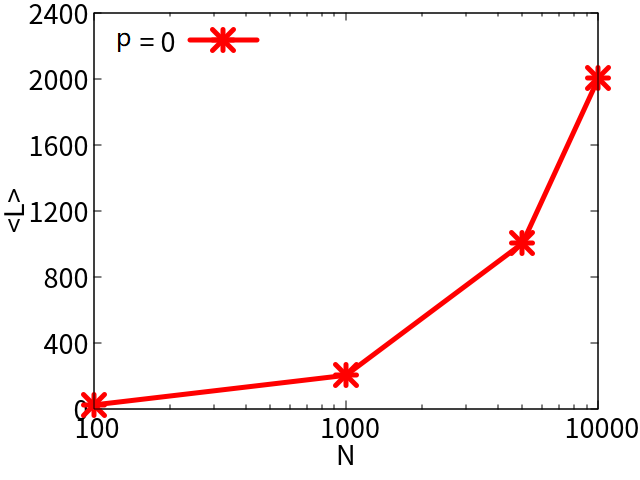}
\end{minipage}%
}%
\subfigure[(b) $p=0$, $m=4$]{
\begin{minipage}[t]{0.5\linewidth}
\centering
\includegraphics[width=60mm]{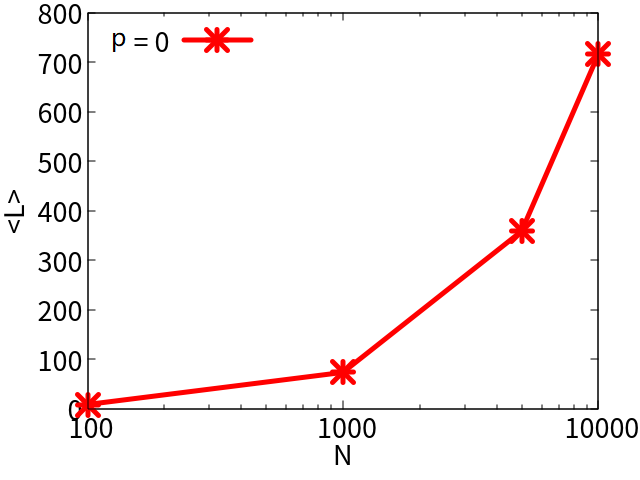}
\end{minipage}%
}%

\subfigure[(c) $p>0$, $m=2$]{
\begin{minipage}[t]{0.5\linewidth}
\centering
\includegraphics[width=60mm]{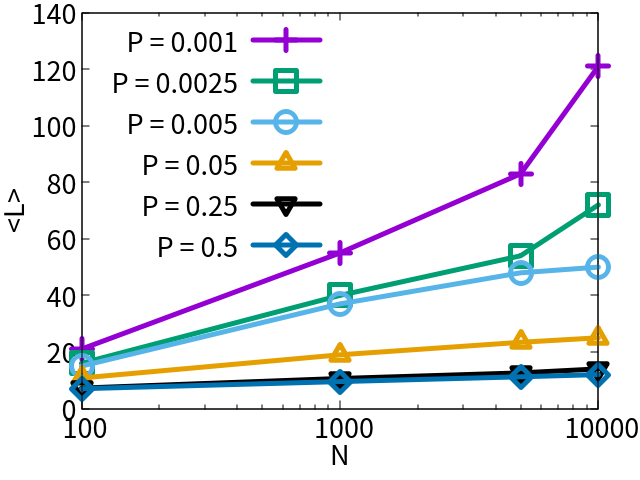}
\end{minipage}%
}%
\subfigure[(d) $p>0$, $m=4$]{
\begin{minipage}[t]{0.5\linewidth}
\centering
\includegraphics[width=60mm]{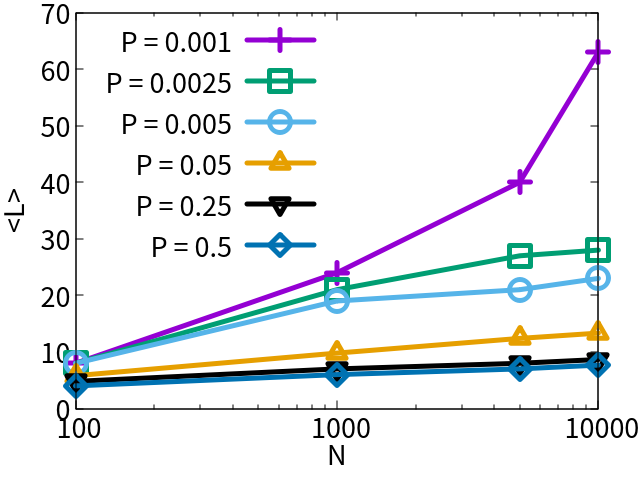}
\end{minipage}%
}%
\centering
\caption{Average path length $\langle L \rangle$ versus node size $N$ at $k^{-\infty}$-attachment. Note that the scale of vertical axis is different in (a)$\sim$(d). Here, $p$ indicates the ratio of random attachment with $m=2$ (right) and $4$ (left). The average path length $\langle L \rangle$ decreases in the cases from $p=0$ to $p=0.05,0.001,0.0025,0.005,0.25$,and $0.5$. When $p$ is larger than a value between $0.0025$ and $0.005$, $\langle L \rangle$ becomes shorter as almost straight lines of $O(\mathrm{log}N)$ in (c) (d). Thus, a small amount of random attachment is necessary for the emergence of the efficient paths: $\langle L \rangle \sim O(\mathrm{log}N)$ in the mixed attachment.}\label{fig7}
\end{figure}


In Fig. \ref{fig5}, green (square mark), brown (pentagon mark), and cyan (circle mark) lines ($\beta=1,4,10$) show that the efficiency trends to increase as the ratio $p$ is larger. We remark that the efficiency rapidly increases for the horizontal axis as shown by black (triangle mark), orange (inverse triangle mark), and red (diamond mark) lines ($\beta=50,100$, and $\beta \to \infty$). This phenomenon corresponds to appearing of chain structures. In comparison with the same color lines or marks, the efficiency in Fig. \ref{fig5} (b) for $m=4$ are higher than that in Fig. \ref{fig5} (a) for $m=2$ by using more links in the emergence of onion-like networks. The average length of the shortest paths is given by $\langle L \rangle \approx 1/E$ in a relation of the arithmetic and the harmonic means. For example, the efficiency $E$ = 0.1, 0.2, and 0.25 correspond to the average path lengths of 10, 5, and 4 hops, respectively. Thus, decreasing efficiency $E$ leads to increasing average path length $\langle L \rangle$. As shown in Fig. \ref{fig7} (a) (c) or (b) (d), the average path length $\langle L \rangle$ decreases in the cases from $p=0$ (top) to $p$ = 0.001, 0.0025, 0.005, 0.05, 0.25, and 0.5 (bottom) denoted by red (asterisk mark), purple (cross mark), green (square mark), cyan (circle mark), orange (triangle mark), black  (inverse triangle mark), and blue (diamond mark) lines, respectively. In particular, rapidly increasing (red) curves are obtained for $p=0$ as the pure $k^{-\infty}$-attachment, therefore these results are not proportional to $\mathrm{log}N$ because of the chain structure as mentioned in the previous subsection. However, the exact critical $p_{c}$ is unknown because of a continuous transition of the average path lengths between $O(N)$ and $O(\mathrm{log}N)$. We estimate that it may be between 0.0025 and 0.005 as shown in Fig. \ref{fig7} (c) (d), in which the average path length becomes longer as almost straight lines of $O(\mathrm{log}N)$ than ones in Fig. \ref{fig7} (a) (b). In addition, the average path length in Fig. \ref{fig7} (d) for $m=4$ is shorter than that in Fig. \ref{fig7} (c) for $m=2$ by using more adding links in comparison with the same color lines or marks. For not only $\beta \to \infty$ but also other values ($\beta=100$), similar results in Fig. \ref{fig7} are obtained including the critical value: $0.0025\leq p_{c}\leq 0.005$. Thus, a small amount of random attachment is necessary for the emergence of the efficient paths in the mixed attachment.\\

\section{Modified p-models with the $k^{-\beta}$-attachment}
In this section, we consider the modified propinquity model (abbreviation as p-model) with the $k^{-\beta}$-attachment. Because the p-model \cite{gallos2019propinquity} is similar to the MED model \cite{hayashi2018new,hayashi2018onion}, a node at the distance $d$ or $\mu+1$ hops from a randomly chosen node is selected as the link destination in both models. We call it distance attachment. The essential difference between MED model and p-model is that the distance attachment is deterministic in MED model but probabilistic in p-model. If there are some nodes in the candidate set of nodes at $d$ or $\mu+1$ hops from a randomly chosen node in the existing nodes, one of them is chosen u.a.r in the original p-model and MED-rand. In spite of the similarity, it is not clear whether p-model can generate onion-like networks. Thus, we modify the random selection to the selection with probability $k^{-\beta}$ in the candidate set in order to investigate the emergence of onion-like networks.\\

\begin{figure}[htbp]
\centering
\subfigure[(a) p-model0]{
\begin{minipage}[t]{0.5\linewidth}
\centering
\includegraphics[width=60mm]{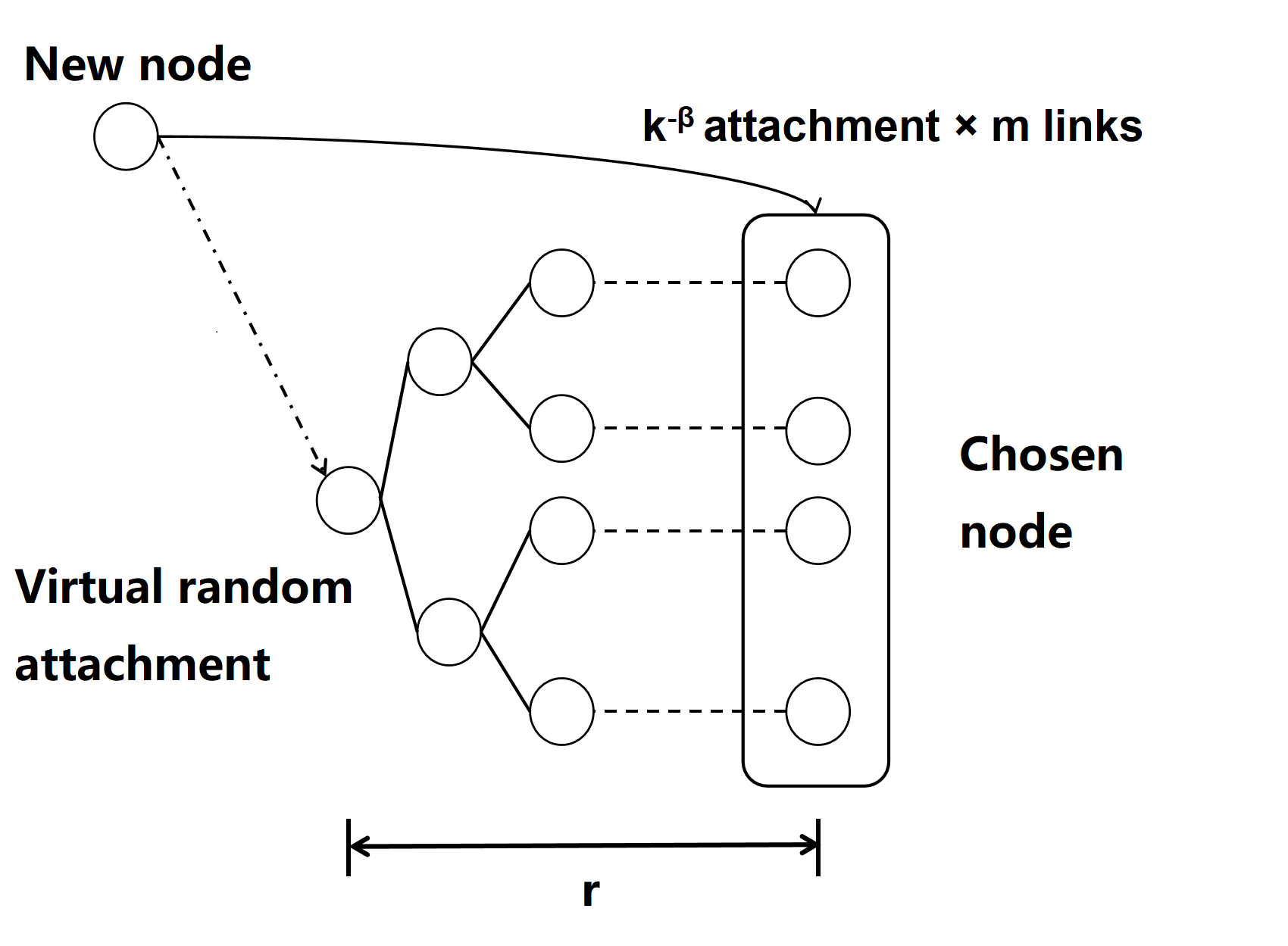}
\end{minipage}%
}%
\subfigure[(b) p-model1]{
\begin{minipage}[t]{0.5\linewidth}
\centering
\includegraphics[width=60mm]{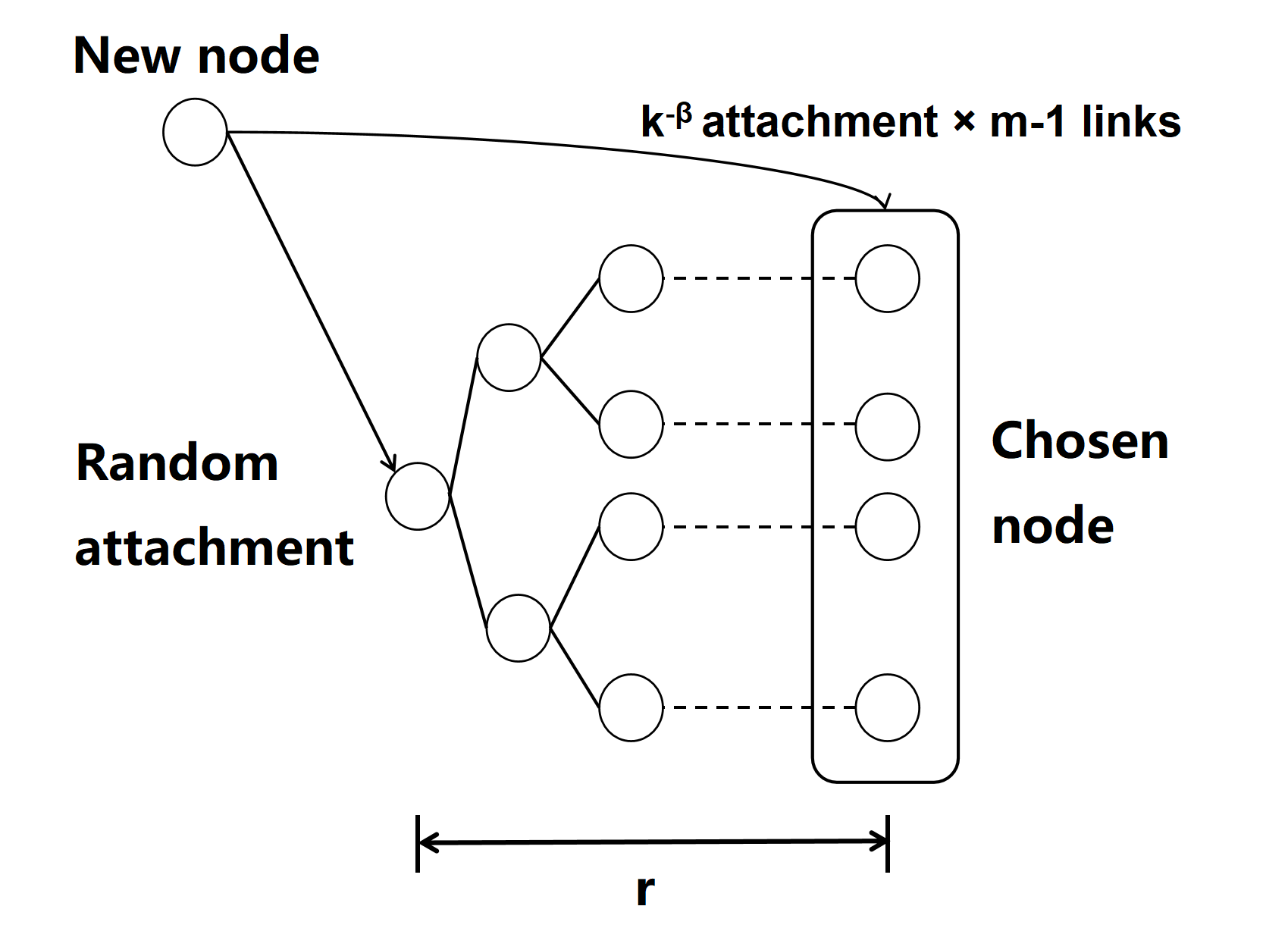}
\end{minipage}%
}%

\subfigure[(c) p-model2]{
\begin{minipage}[t]{0.5\linewidth}
\centering
\includegraphics[width=60mm]{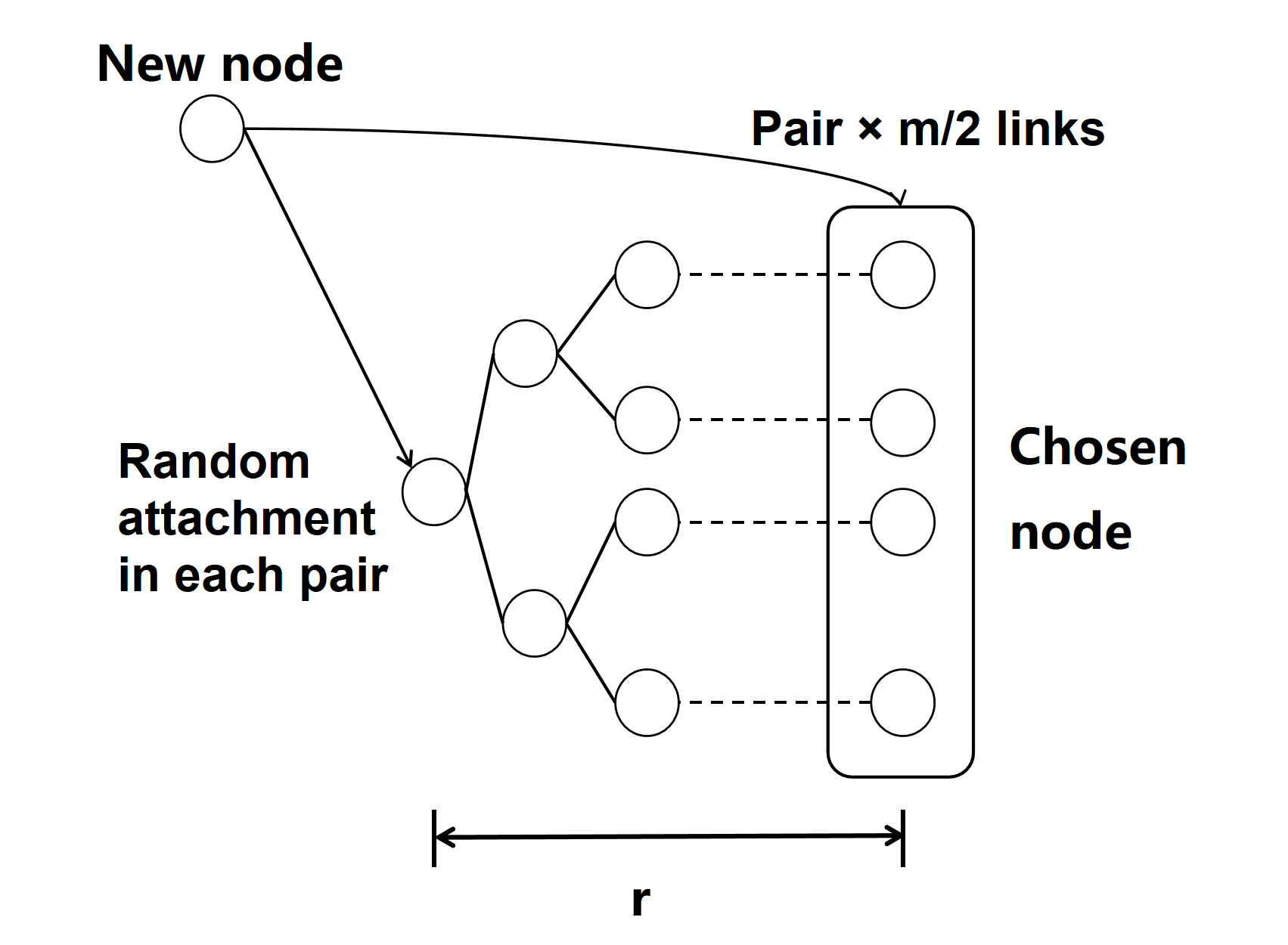}
\end{minipage}%
}%
\subfigure[(d) M-MED model]{
\begin{minipage}[t]{0.5\linewidth}
\centering
\includegraphics[width=60mm]{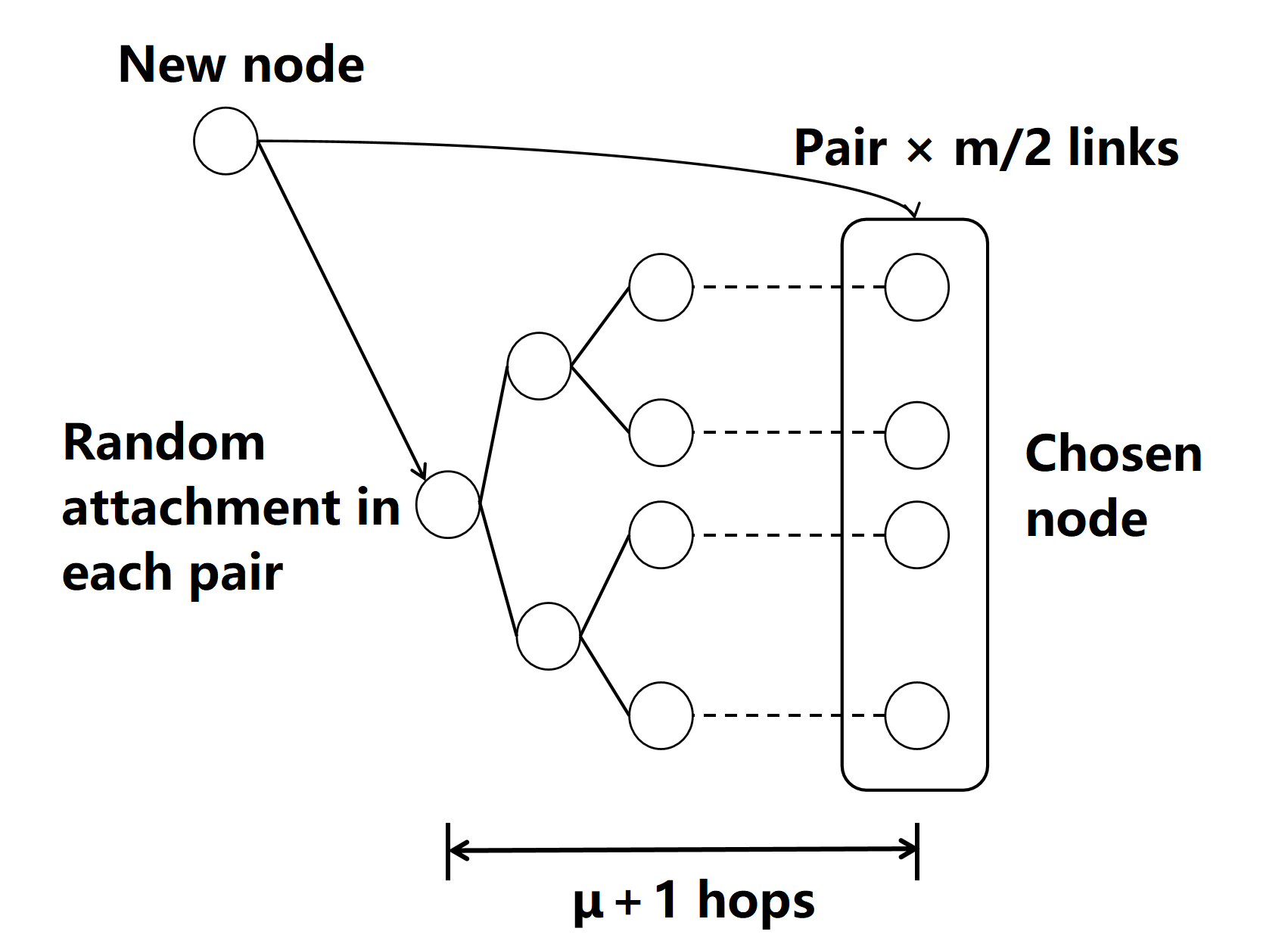}
\end{minipage}%
}%
\centering
\caption{Growing process in p-models and M-MED model. As shown in (a)$ \sim $(c), modified p-models with different number of random attachments \cite{gallos2019propinquity}. As shown in (d), M-MED model is modified with the $k^{-\beta}$-attachment from MED model \cite{hayashi2018new,hayashi2018onion} instead of the minimum degree selections. Distances $d$ and $\mu+1$ are measured by the number of hops from a randomly chosen node in p-models and MED model.}\label{fig15}
\end{figure}

In the modification of p-model, the attached node $i$ is chosen with double probabilities according to its degree $k_{i}$ and a distance $d$. We consider only the case of $m=4$ links per time step, because an onion-like network emerges for $m=4$ as shown previously. Fig. \ref{fig15} illustrates the growing process in p-model0, p-model1, p-model2 and M-MED model. Here, the suffix numbers 0, 1, and 2 mean no random attachment but virtual selection, one random attachment, and pair of random and distance (or intermediation) attachments, respectively. M-MED model is introduced as a modification of MED model with the $k^{-\beta}$-attachment in order to compare with p-model2 based on similar $m/2$ pairs of attachments. $\alpha>0$, $\beta>0$, and $d=\mu+1\geq 1$ are given parameters. 

\begin{description}
\item[p-model0] A node is selected u.a.r in the existing nodes. For each of $m$ links from a new node, there are two step selections as follows. A candidate set of nodes at a distance $d$ from the virtually selected node is chosen with probability proportional to $d^{-\alpha}$. Then a node $i$ is chosen with probability proportional to $k_{i}^{-\beta}$ in the candidate set. Of course, when there is only one node in the candidate set, it is chosen. 
\item[p-model1] The first attached node is chosen u.a.r in the existing nodes. For each of other $m-1$ links, the above process except virtual selection in p-model0 is repeated as similar to the original p-model \cite{gallos2019propinquity}, but with the modification by the $k^{-\beta}$-attachment in the candidate set. 
\item[p-model2] In each of $m/2$ pairs of random and intermediation attachments, one of link destination is chosen u.a.r in the existing nodes. A candidate set of nodes at a distance $d$ from the randomly chosen pair node is chosen with probability proportional to $d^{-\alpha}$. Another node $i$ is chosen with probability proportional to $k_{i}^{-\beta}$ in the candidate set for each pair. 
\item[M-MED model] As similar to p-model2, $m/2$ pair nodes are chosen, but the candidate set of nodes at a constant distance $\mu+1$ is determined for each randomly chosen pair node. The modification denoted by M- means applying of the $k^{-\beta}$-attachment instead of random selection in the candidate set.  
\end{description}

\begin{figure}[tbp]
\centering
\subfigure[(a) Assortativity, p-model0]{
\begin{minipage}[t]{0.5\linewidth}
\centering
\includegraphics[width=50mm]{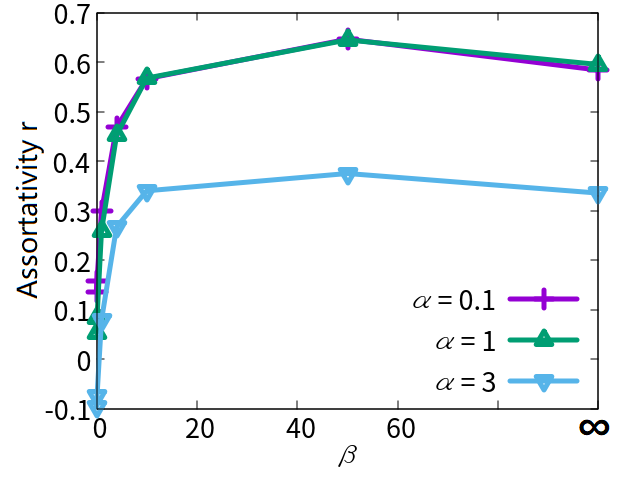}
\end{minipage}%
}%
\subfigure[(d) Assortativity, p-model1]{
\begin{minipage}[t]{0.5\linewidth}
\centering
\includegraphics[width=50mm]{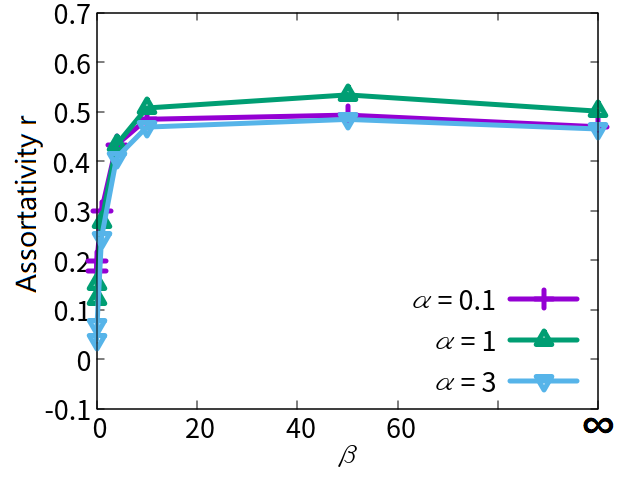}
\end{minipage}%
}%

\subfigure[(b) Robustness, p-model0]{
\begin{minipage}[t]{0.5\linewidth}
\centering
\includegraphics[width=50mm]{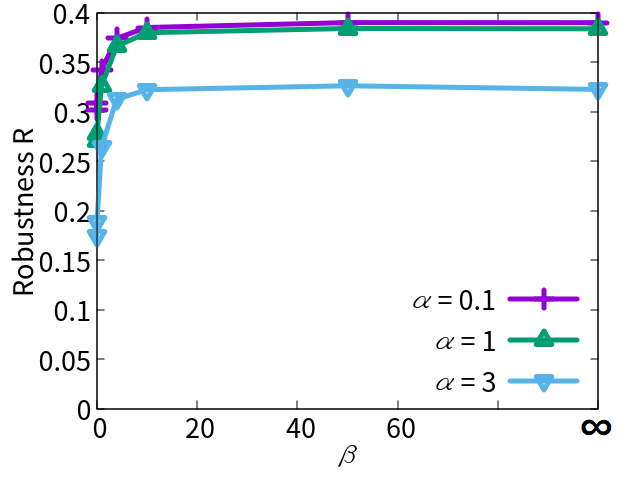}
\end{minipage}%
}%
\subfigure[(e) Robustness, p-model1]{
\begin{minipage}[t]{0.5\linewidth}
\centering
\includegraphics[width=50mm]{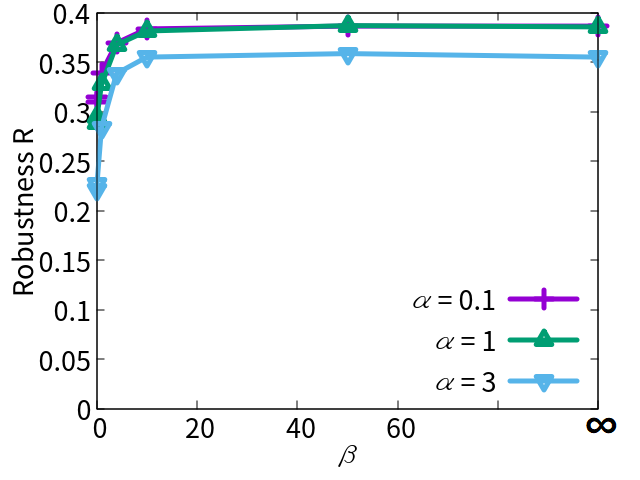}
\end{minipage}%
}%

\subfigure[(c) Efficiency, p-model0]{
\begin{minipage}[t]{0.5\linewidth}
\centering
\includegraphics[width=50mm]{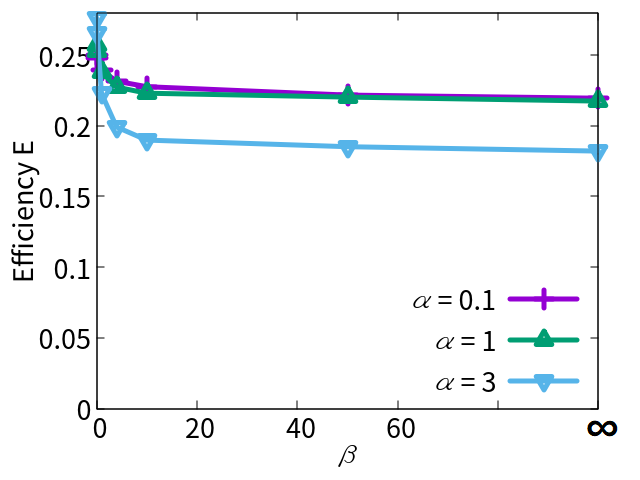}
\end{minipage}%
}%
\subfigure[(f) Efficiency, p-model1]{
\begin{minipage}[t]{0.5\linewidth}
\centering
\includegraphics[width=50mm]{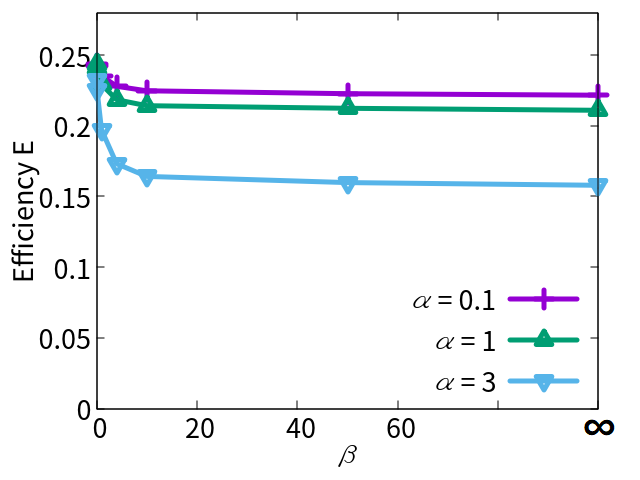}
\end{minipage}%
}%

\centering
\caption{Three basic measures $r$ (top), $R$ (medium), $E$ (bottom) versus parameter $\beta$ in the p-model0 and p-model1 at size $N=5000$ with $m$=4 links per time step. As shown in (a) (d), assortativity increase rapidly in p-model0 and p-model1 for $\alpha=0.1,1,3$ (purple, green and cyan lines with cross, triangle, and inverse triangle marks) as $\beta$ is from $0$ to $10$, and becomes steady after $\beta>10$. As shown in (b) (e), robustness increases rapidly in $\beta<10$. These networks have both high assortativity and robustness as onion-like structure. As shown in (c) (f), efficiency slightly decreases for $\alpha=0.1,1,3$ as $\beta$ is larger from $0$ to $10$, however, efficiency becomes steady after $\beta>10$. \textbf{}}\label{fig11}
\end{figure}

In the following discussions, we compare with the p-model0 and p-model1 to investigate the effect of different number of random attachments on the emergence of onion-like networks. Then, we similarly compare with the p-model2 and M-MED to investigate the effect of distance $d=\mu+1$ on the emergence of onion-like networks. Fig. \ref{fig11} (a) shows that the assortativity increase rapidly in p-model0 for $\alpha$ = 0.1 and 1 (purple and green lines with cross and triangle marks) as $\beta$ grows from $0$ to $10$, and becomes steady after $\beta>10$. While the case of $\alpha=3$ (cyan line with inverse triangle marks) as almost connecting to the nearest neighbors of a randomly chosen node have weak assortativity. This case corresponds to the Duplication model \cite{pastor2003evolving} which generates a SF network \cite{yang2013scale} by slightly different rules: always connecting to the nearest neighbors of a virtually random selection node for a varying number (not constant $m$) of links. Moreover, the assortativity for $\alpha$ = 0.1 and 1 (purple and green lines with cross and triangle marks) in p-model1 are lower than ones in p-model0 as shown in Fig. \ref{fig11} (a) (d). There is no significant difference between $\alpha=3$ (cyan line with inverse triangle marks)  and $\alpha$ = 0.1 and 1 (purple and green lines with cross and triangle marks) in Fig. \ref{fig11} (d). Therefore, the effect of distance $d$ is not dominant for increasing the assortativity. Furthermore, robustness increases rapidly with $\beta$ for the case of $\alpha$ = 0.1 and 1 (purple and green lines with cross and triangle marks) in $\beta<10$ as shown in Fig. \ref{fig11} (b), while the case of $\alpha=3$ (cyan line with inverse triangle marks) have slightly weak robustness. These networks have high assortativity and robustness as onion-like structure.\\

\begin{figure}[htbp]
\centering

\subfigure[(a) Assortativity, p-model2]{
\begin{minipage}[t]{0.5\linewidth}
\centering
\includegraphics[width=50mm]{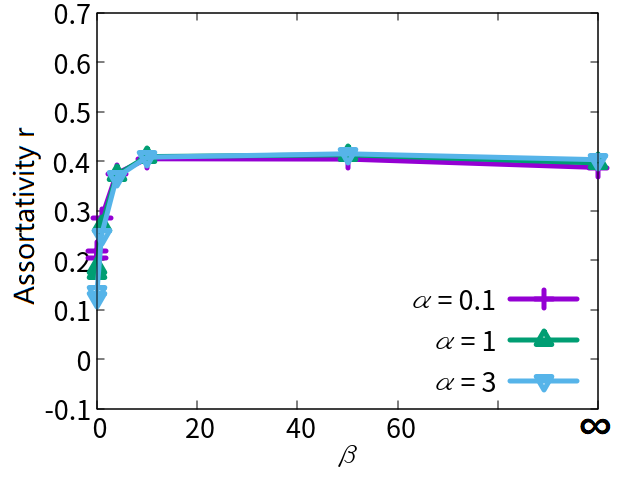}
\end{minipage}%
}%
\subfigure[(d) Assortativity, M-MED]{
\begin{minipage}[t]{0.5\linewidth}
\centering
\includegraphics[width=50mm]{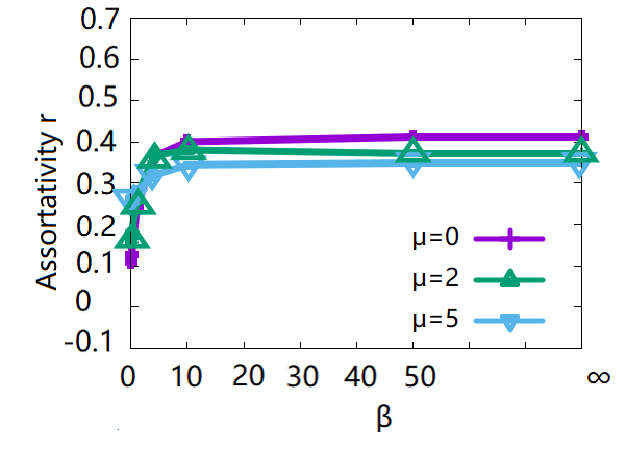}
\end{minipage}%
}%

\subfigure[(b) Robustness, p-model2]{
\begin{minipage}[t]{0.5\linewidth}
\centering
\includegraphics[width=50mm]{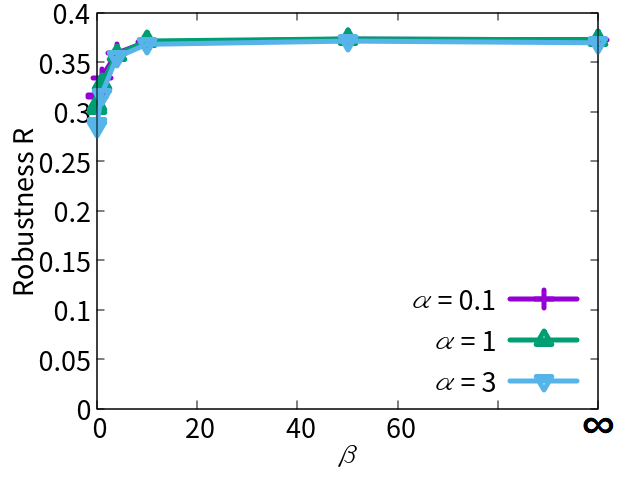}
\end{minipage}%
}%
\subfigure[(e) Robustness, M-MED]{
\begin{minipage}[t]{0.5\linewidth}
\centering
\includegraphics[width=50mm]{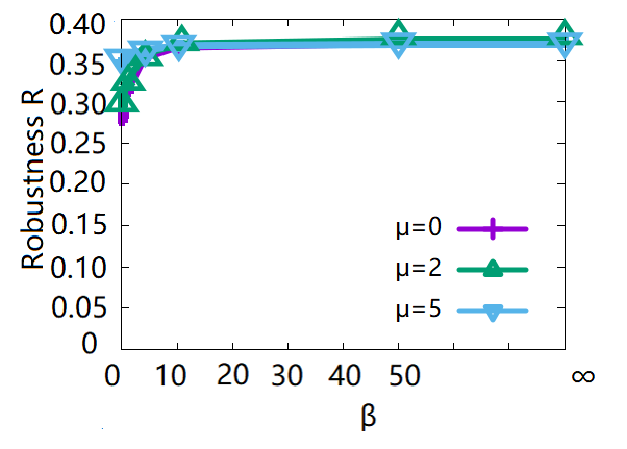}
\end{minipage}%
}%

\subfigure[(c) Efficiency, p-model2]{
\begin{minipage}[t]{0.5\linewidth}
\centering
\includegraphics[width=50mm]{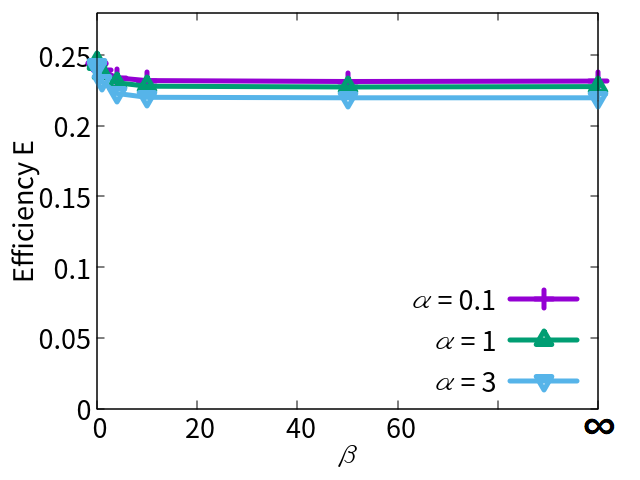}
\end{minipage}%
}%
\subfigure[(f) Efficiency, M-MED]{
\begin{minipage}[t]{0.5\linewidth}
\centering
\includegraphics[width=50mm]{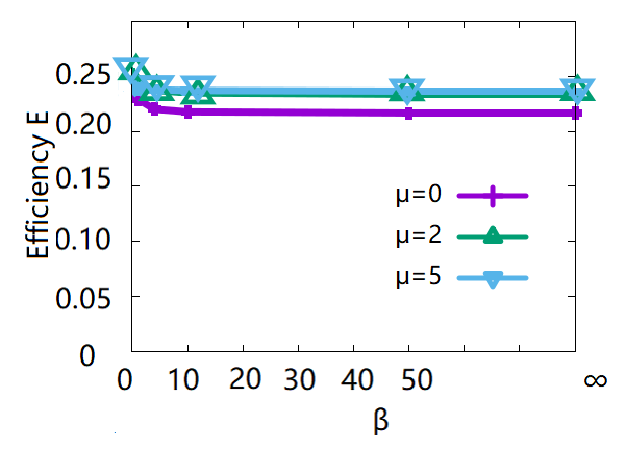}
\end{minipage}%
}%
\centering
\caption{Three basic measures $r$, $R$, $E$ versus parameter $\beta$ in the p-model2 (top) and M-MED model (bottom) at size $N=5000$ with $m$=4 links per time step. As shown in (a) (d), assortativity increase rapidly in p-model2 and M-MED model for $\alpha=0.1,1,3$ and $\mu=0,2,5$ (purple, green and cyan lines with cross, triangle, and inverse triangle marks) as $\beta$ is from $0$ to $10$, and becomes steady after $\beta>10$. As shown in (b) (e), robustness increases rapidly in $\beta<10$. As shown in (c) (f), the efficiency are similarly high in comparison with the same color lines or marks. Moreover, no significant difference between cyan (inverse triangle marks), purple (cross mark), and green (triangle mark) lines shows that the effect of distance $d=\mu+1$ is not dominant for increasing the assortativity, robustness, and efficiency. \textbf{}}\label{fig13}
\end{figure}

Next, we discuss the properties for the assortativity as degree-degree correlations $r$, the robustness index $R$, and the efficiency $E$ in both p-model2 and M-MED model. Here, $\alpha=3$ or $\mu=0$, $\alpha=1$ or $\mu=2$, and $\alpha=0.1$ or $\mu=5$ correspond to local, middle, and far intermediation, respectively. Fig. \ref{fig13} (a) (d) for p-model2 and M-MED model show that the assortativity for local , middle, and far intermediation are similarly high in comparison with the same color lines or marks. Moreover, there is no significant difference between p-model2 and M-MED model. We should notice that the distance $d=\mu+1$ does not so much affect assortativity, robustness, and efficiency. Table 3 summarizes the emergence of onion-like networks for $\beta$ value in modified p-models and MED model. When $\beta>0$, all modified models become onion-like structure.\\

\begin{figure}[htbp]
\centering
\subfigure[(a) p-model0]{
\begin{minipage}[t]{0.5\linewidth}
\centering
\includegraphics[width=60mm]{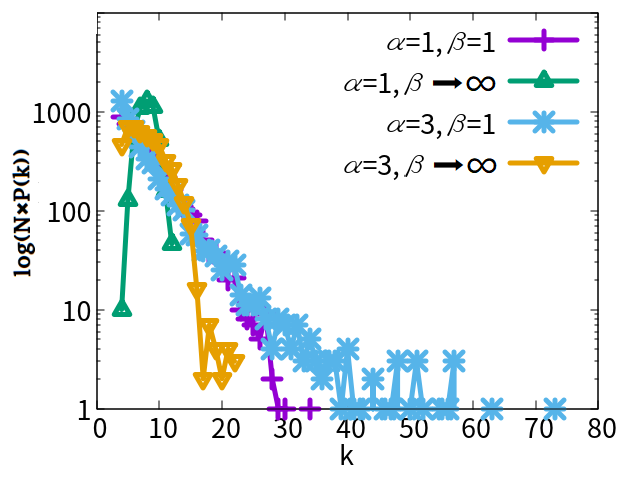}
\end{minipage}%
}%
\subfigure[(b) p-model1]{
\begin{minipage}[t]{0.5\linewidth}
\centering
\includegraphics[width=60mm]{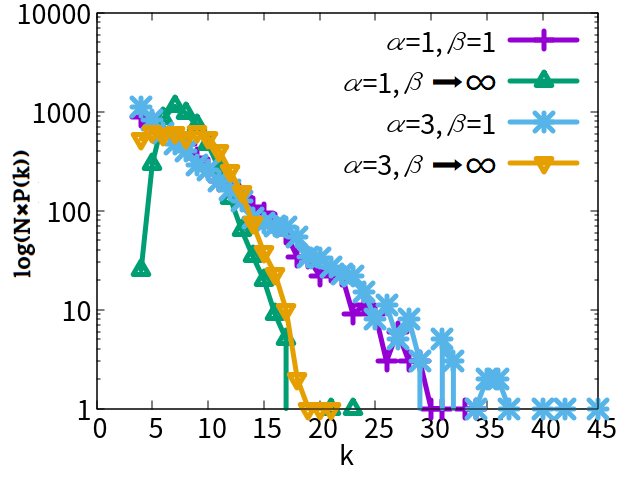}
\end{minipage}%
}%

\subfigure[(c) p-model2]{
\begin{minipage}[t]{0.5\linewidth}
\centering
\includegraphics[width=60mm]{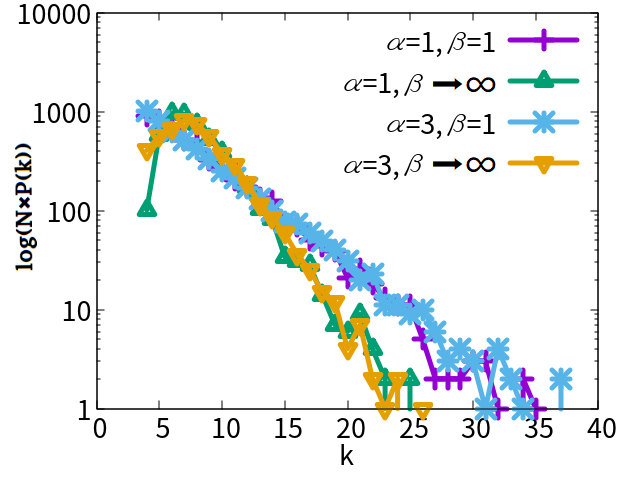}
\end{minipage}%
}%
\subfigure[(d) M-MED]{
\begin{minipage}[t]{0.5\linewidth}
\centering
\includegraphics[width=60mm]{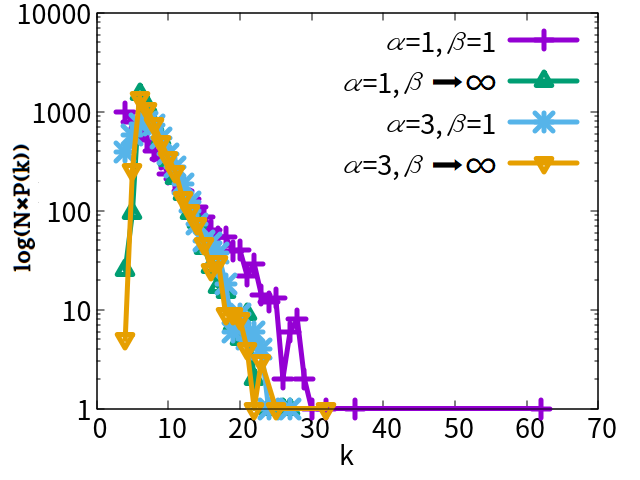}
\end{minipage}%
}%
\centering
\caption{The degree distribution in four modified models in growing networks with $m=4$. For each of (a) (b) (c) (d), there are similar distributions in p-model0, p-model1, p-model2, and M-MED model. The tails of distributions are linear in a semi-logarithmic plot as exponential distributions. The slopes of green (triangle mark) and orange (inverse triangle marks) lines for $\beta \to \infty$ are steeper than ones of the purple (cross mark) and cyan (asterisk mark) lines for $\beta=1$. Thus, large degrees are bounded as $\beta$ increases. In other words, no hubs exist.}\label{fig14}
\end{figure}

In addition, Fig. \ref{fig14} (a), (b), (c), and (d) show the degree distribution in p-model0, p-model1, p-model2, and M-MED model. The tails of distributions are linear in a semi-logarithmic plot as exponential distributions. The slopes of green (triangle mark) and orange (inverse triangle mark) lines for $\beta \to \infty$ are steeper than ones of the purple (cross mark) and cyan (asterisk mark) lines for $\beta=1$. Thus, large degrees are bounded as $\beta$ increases, it is considered to be strong robustness because of no huge hubs.  
\begin{table}[htbp]
\centering
\begin{tabular}{c|c|c|c|c}
 & RA: $\beta=0$ & $0<\beta<\infty$ & kmin: $\beta \to \infty$ & Ratio of RA \\ \hline
p-model0 & \tabincell{c}{Not onion-like \\(D-D: SF at $\alpha \to \infty$) }  & Onion-like & Onion-like & 0 \\ \hline
p-model1 & Not onion-like & Onion-like & Onion-like & 1/m  \\ \hline
p-model2 & Not onion-like & Onion-like & Onion-like & 0.5  \\ \hline
M-MED model & \tabincell{c}{Not onion-like \\(MED-rand) } & Onion-like & \tabincell{c}{Onion-like \\(MED-kmin) } & 0.5  \\ \hline
\end{tabular}
\caption{Emergence of onion-like networks for $\beta$ value in the modified p-models and MED model. MED-rand and MED-kmin are special cases at $\beta=0$ and $\infty$ \cite{hayashi2018new,hayashi2018onion}. RA and D-D denote random attachment and duplication divergence models \cite{pastor2003evolving}, respectively.}\label{table3}
\end{table}

In comparison with the same color lines or marks, the efficiency in (b) for $m=4$ is higher than that in (a) for $m=2$ by using more adding links in the emergence of onion-like networks.

\section{Conclusion}
In summary, to make a strongly robust onion-like network with positive assortativity, we have studied the $k^{-\beta}$-attachment which interpolates random and the minimum degree attachments by a parameter $\beta\geq0$. In particular, we show that the $k^{-\beta}$-attachment generates onion-like networks with both high assortativity and robustness for a large $\beta>0$. However, a chain structure is obtained with the inefficient $O(N)$ longest paths (as the diameter) at $\beta \to \infty$. Thus, we have considered a mixed attachment of random and the $k^{-\beta}$ attachments to get the efficient paths as the average path length of $O(\mathrm{log}N)$. We have found that a small amount of random attachment is necessary for the emergence of the efficient paths in the mixed attachment. In addition, we modify p-model \cite{gallos2019propinquity} and MED model \cite{hayashi2018new,hayashi2018onion} with the $k^{-\beta}$-attachment in order to investigate the effects of the $k^{-\beta}$-attachment and distance $d=\mu+1$ on the emergence of onion-like networks. The obtained results for assortativity as the degree degree correlations, robustness, and efficiency show that the onion-like network can be generated by the $k^{-\beta}$-attachment in a wide range of $\beta>0$, and that the distance from a randomly chosen node does not so much affect the emergence. These results will contribute to a better understanding of generating onion-like networks. Instead of rich get richer rule in selfish preferential attachment, a small amount of random attachment gives an encountering link to a node equally. While the minimum degree attachment gives a helpful chance of linking to a poor and unlikely useful node with a small degree, it leads to increasing the tolerance of connectivity against attacks in the network. Such explanations may be useful for realizing a better network with both strong robustness and high efficiency in future systems.

\section*{Acknowledgements}
We would like to express our thanks to anonymous reviewers for their valuable comments in improving the readability of this manuscript. This research is supported in part by JSPS KAKENHI Grant Number JP.21H03425.

\bibliography{mybibfile}

\end{document}